\documentclass[a4paper,11pt]{article}
\pdfoutput=1 

\usepackage{jcappub} 

\usepackage[T1]{fontenc} 

\usepackage{aas_macros}
\usepackage{natbib}
\usepackage{amsfonts}
\usepackage{amsmath}
\usepackage{amssymb}
\usepackage{graphicx}
\usepackage{color}
\usepackage{hyperref}

\defcitealias{ps22}{PS22}
\defcitealias{ps23}{PS23}

\newcommand{\p}{\ensuremath{\partial}}

\newcommand{\xx}{\ensuremath{\mathbf{x}}}

\newcommand{\Mpch}{\ensuremath{h^{-1}{\rm Mpc}}}

\newcommand{\avg}[1]{\ensuremath{\left\langle \,#1\, \right\rangle}}
\newcommand{\e}[1]{\ensuremath{{\rm e}^{#1}}}

\newcommand{\der}{\ensuremath{{\rm d}}}

\newcommand{\eqn}[1]{equation~\eqref{#1}}

\newcommand{\beq}{\begin{equation}}
\newcommand{\eeq}{\end{equation}}
\newcommand{\Cal}[1]{\ensuremath{\mathcal{#1}}}

\newcommand{\biseq}{\texttt{BiSequential}}

\newcommand{\xilin}[1]{\ensuremath{\xi_{\rm lin}(#1)}}
\renewcommand{\ss}{\ensuremath{\mathbf{s}}}
\newcommand{\rr}{\ensuremath{\mathbf{r}}}

\newcommand{\bt}{\ensuremath{\boldsymbol{\theta}}}
\newcommand{\bb}{\ensuremath{\mathbf{b}}}
\newcommand{\ww}{\ensuremath{\mathbf{w}}}

\title{\boldmath Model-agnostic basis functions for the 2-point correlation function of dark matter in linear theory}

\author[a]{Aseem Paranjape}
\author[b,c]{and Ravi K. Sheth}

\affiliation[a]{Inter-University Centre for Astronomy \& Astrophysics,\\ Ganeshkhind, Post Bag 4, Pune 411007, India}
\affiliation[b]{Center for Particle Cosmology, University of Pennsylvania,\\ 209 S. 33rd St., Philadelphia, PA 19104, USA}

\affiliation[c]{The Abdus Salam International Center for Theoretical Physics,\\ Strada Costiera, 11, Trieste 34151, Italy}
 
\emailAdd{aseem@iucaa.in}
\emailAdd{shethrk@physics.upenn.edu}

\abstract{
We consider approximating the linearly evolved 2-point correlation function (2pcf) of dark matter \xilin{r;\bt} in a cosmological model with parameters \bt\ as the linear combination $\xilin{r;\bt}\approx\sum_i\,b_i(r)\,w_i(\bt)$, where the functions $\Cal{B}=\{b_i(r)\}$ form a \emph{model-agnostic basis} for the linear 2pcf. 
This decomposition is important for model-agnostic analyses of the baryon acoustic oscillation (BAO) feature in the nonlinear 2pcf of galaxies that fix \Cal{B} and leave the coefficients $\{w_i\}$ free. 
To date, such analyses have made simple but sub-optimal choices for \Cal{B}, such as monomials.
We develop a machine learning  framework for systematically discovering a \emph{minimal} basis \Cal{B} that describes \xilin{r} near the BAO feature in a wide class of cosmological models. 
We use a custom architecture, denoted \biseq, for a neural network (NN) that explicitly realizes the separation between $r$ and \bt\ above. 
The optimal NN trained on data in which only $\{\Omega_{\rm m},h\}$ are varied in a \emph{flat} $\Lambda$CDM model produces a basis \Cal{B} comprising $9$ functions capable of describing \xilin{r} to $\sim0.6\%$ accuracy in \emph{curved} $w$CDM models varying 7 parameters within $\sim5\%$ of their fiducial, flat $\Lambda$CDM values. 
Scales such as the peak, linear point and zero-crossing of \xilin{r} are also recovered with very high accuracy. 
We compare our approach to other compression schemes in the literature, and speculate that \Cal{B} may also encompass \xilin{r} in modified gravity models near our fiducial $\Lambda$CDM model. 
Replacing the \emph{ad hoc} bases in model-agnostic BAO analyses with our basis functions can potentially lead to significant gains in constraining power. 
}


\begin{document}
\maketitle
\flushbottom

\section{Introduction}
\label{sec:intro}
The baryon acoustic oscillation (BAO) feature in the $2$-point clustering of tracers such as galaxies is a key science driver of a number of current and upcoming large-volume surveys \cite{alam+21,DESI,DESI2024-cosmo}. The location and shape of the BAO feature is known to contain a wealth of information regarding, both, fundamental cosmology and the nonlinear evolution of matter in the Universe \cite{eisenstein+05,cs08}. Extracting the information on fundamental (or primordial) cosmological variables and cleanly disentangling it from the effects of nonlinear growth is a major avenue of current research in cosmology. Traditional analyses of the BAO feature for cosmology have typically relied on using model-dependent templates within, say, the Lambda-cold dark matter ($\Lambda$CDM) framework to extract cosmological information \cite{cuesta+16,beutler+17,blomqvist+19,dumasdesbourbouz+20,gil-marin+20,extractor2020,abbott+22}. More recently, there has also been a thrust towards developing \emph{model-agnostic} analysis frameworks that instead combine generic descriptions of the clustering of dark matter in linear theory with minimalistic assumptions regarding nonlinear evolution (such as the Zel'dovich approximation) to describe the 2-point correlation function (2pcf) of nonlinearly evolved tracers \cite{LP2016,LPboss,nsz21a,nsz21b,nsz22,ps22,hzs23}. While this early work focused on the monopole $\xi^{(0)}$ of the nonlinear 2pcf, more recent work \citep{ps23} has extended this framework to all relevant redshift space multipoles $\xi^{(\ell)}$, $\ell=0,2,4$ expected to be accessible in upcoming surveys such as that of the Dark Energy Spectroscopic Instrument \citep[DESI;][]{DESI,DESI2024-cosmo}.

A key building block in a model-agnostic treatment of the BAO feature for cosmological inference is the use of a set of basis functions for describing the shape of the real space dark matter 2pcf \xilin{r}\ in linear theory.\footnote{Note that \xilin{r} is a purely theoretical quantity, to be distinguished from, say, the observable nonlinear 2pcf monopole $\xi^{(0)}$ mentioned above.} Previous studies \cite{nsz21b,ps22,ps23} have used basis sets comprised of simple functions such as mononomials $\{1,r,r^2,\ldots\}$, or their linear combinations based on PCA considerations. The choice of monomials as fundamental building blocks is arbitrary, however, and the use of PCA necessarily requires additional knowledge of the eventual data set one is targeting, along with its covariance matrix. It is generically also true that polynomial, or even PCA, bases are typically over-complete when used for inference with BAO multipoles in redshift space, leading to a leakage of information from cosmologically interesting parameters into the redundant degrees of freedom of the chosen basis.

It is therefore interesting to ask whether one can discover a minimal basis set of functions to accurately describe the shape of \xilin{r}\ over a wide class of cosmological models. Some recent work in this direction has promoted the use of a basis set based on the expected shape of the BAO feature in $\Lambda$CDM, along with the broadband shape of the 2pcf surrounding it \cite{s-fh21}. A broader version of this question is now a routine part of machine learning (ML) approaches to function approximation. A typical ML solution to approximating a non-linear, multi-dimensional function $f(\xx)$ where $\xx\in \mathbb{R}^n$ would consist of training a deep neural network with, say, $L$ hidden layers followed by an output layer. The universal approximation theorem \citep[e.g.,][]{deeplearning-gbc16} guarantees the existence of a suitable network that can approximate the function with any desired choice of accuracy. The last of these hidden layers (of dimension $n_L$, say) provides an $n_L$-dimensional representation $\{b_i(\xx)\}_{i=1}^{n_L}$ of the input \xx, such that a linear combination of the functions $b_i$ approximates the target $f(\xx)$ (modulo a last, possibly non-linear activation). In other words, the ML approach using a neural network (NN) constructs a basis as the output of its last hidden layer (i.e., the penultimate layer of the NN). 

In our case, the target function is \xilin{r;\bt}, where $r$ is the scalar magnitude of the separation vector and \bt\ is an $n_{\rm p}$-dimensional vector of cosmological parameters. For the discussion below, we will focus on curved $w$CDM cosmologies with \bt\ being a subset of $\{\Omega_{\rm m},h,n_{\rm s},A_{\rm s},\Omega_{\rm b},\Omega_{\rm k},w_0\}$, i.e., $n_{\rm p}\leq7$. We will comment later on cosmologies based on modified gravity. Our goal is to discover a minimal basis set $\Cal{B} = \{b_i(r)\}_{i=1}^{n_\ast}$, comprised of $n_\ast$ one-dimensional functions, such that linear combinations of the $b_i$ can accurately approximate \xilin{r;\bt} for \emph{any} choice of \bt\ in some chosen range. In other words, we wish to approximate 
\beq
\xilin{r;\bt} \approx \sum_{i=1}^{n_\ast}\,b_i(r)\,w_i(\bt) \equiv \bb(r)\cdot\ww(\bt)\,,
\label{eq:funcapprox}
\eeq
for some appropriate choice of weights $\Cal{W}=\{w_i(\bt)\}$. Effectively, this would separate the information contained in the specific choice of \bt\ into the weights \Cal{W} while isolating the common shape degrees of freedom into the basis set \Cal{B}. This separation will be important later for the capacity of the description to generalize to other classes of cosmology.
Similar questions have been recently addressed in the context of the linear power spectrum (i.e., the Fourier transform of \xilin{r}) and other Fourier space functions \citep{aaz21,philcox+21,eggemeier+23,bcv24}, using techniques based on singular value decomposition (SVD) of matrices constructed by evaluating the relevant function on a fixed grid of (Fourier) scales and a fixed set of `template' cosmological models. We will, however, use deep NNs with a custom architecture for achieving the separability mentioned above.

The paper is organized as follows. In section~\ref{sec:data}, we first describe our choices of fiducial cosmology and the ranges of parameter variations around it, along with the chosen range of separations we will model. Next, we describe the construction of the training, validation and test data sets used in our later ML analysis. In section~\ref{sec:neuralnetwork}, we motivate and describe the custom network architecture used in this work, followed by details of the procedures we use for training, validation and testing of the selected NN. Section~\ref{sec:results} presents our results, followed by a discussion in section~\ref{sec:discuss} that highlights some salient features of our analysis and places our results in the context of other approaches in the literature. We conclude in section~\ref{sec:conclude}.

\section{Data set}
\label{sec:data}
The ML approach requires the construction of training, validation and test data sets, which we describe here. There are two conceivable approaches to constructing a training data set for learning the function \xilin{r;\bt}, which in turn affect the architecture of the NN. In the first, one would fix a vector \rr\ of, say, $n_r$ values of $r$ in some range $[r_{\rm min},r_{\rm max}]$ once and for all and treat the target output as an $n_r$-dimensional vector \xilin{\rr;\bt}, with the input being only \bt. In the second, one would treat both $r$ and \bt\ as inputs -- so that the overall input is $(n_{\rm p}+1)$-dimensional -- and treat the output \xilin{r;\bt} as a scalar. The first approach (common in cosmological emulation literature, \cite{aaz21,eggemeier+23,bcv24}) has the advantage of being amenable to matrix treatments and SVD, but does not easily allow for generalizability to arbitrary $r\in[r_{\rm min},r_{\rm max}]$ while maintaining accuracy. E.g., in this approach one typically discards all but the smoothest eigenvectors obtained from the SVD, so as to keep the emulation problem using NNs tractable. For evaluating the 2pcf at a value of $r$ not coinciding with one of the components of \rr, one would then typically need additional interpolations, especially when estimating scales such the linear point or zero-crossing with high precision. Since we are eventually interested in computing such scales along with accurate weighted integrals of \xilin{r;\bt},
we adopt the second approach which allows for evaluations at arbitrary $r\in[r_{\rm min},r_{\rm max}]$ from the outset, for an appropriate choice of training set.

\subsection{Cosmology and range}
\label{subsec:cosmo}
Our fiducial cosmology is a flat $\Lambda$CDM model with parameters $\{\Omega_{\rm m} = 0.3153,h = 0.6737,\ln(10^{10}A_{\rm s}) = 3.045,n_{\rm s} = 0.9649,\Omega_{\rm b} = 0.04929,\Omega_{\rm k} = 0.0,w_0 = -1.0\}$ 
consistent with the results of \cite{Planck18-VI-cosmoparam}. We use the publicly available code \textsc{class} \citep{class-I,class-II}\footnote{\url{http://class-code.net/}} to compute the linear power spectrum $P_{\rm lin}(k;\bt)$ at $z=0$ and numerically integrate over $k$ to obtain the ground truth for \xilin{r;\bt} using
\beq
\xilin{r;\bt} = \int\frac{\der k}{k}\,\frac{k^3P_{\rm lin}(k;\bt)}{2\pi^2}\,\frac{\sin(kr)}{kr}\,.
\eeq
Throughout, we choose to work with separations $r$ in units of \Mpch\ with $h$ set to its fiducial value. \emph{Below, whenever no confusion can arise, we will use $h$ to denote this fiducial value.} When varying $h$ in the evaluation of the 2pcf to, say, $h_\ast$, we appropriately rescale the argument of the 2pcf and evaluate $\xilin{rh_\ast/h;\bt}$, while consistently requiring \textsc{class} to use $k$ in units of $h_\ast{\rm Mpc}^{-1}$.

\begin{figure}[h!]
\centering
\includegraphics[width=0.45\textwidth]{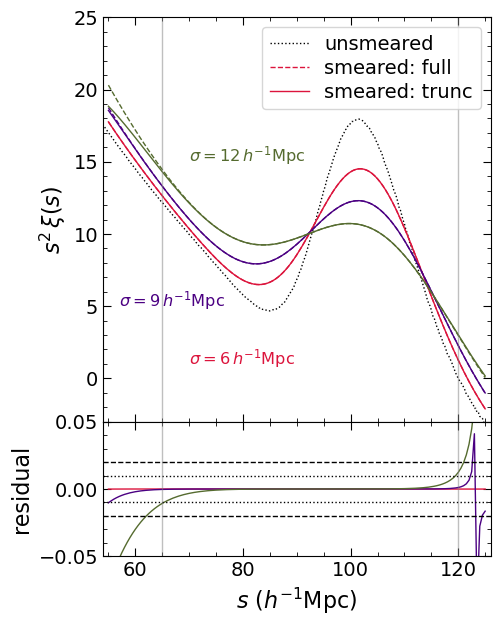}
\caption{Effect of truncation of integral over $r$ to the range $r\in[30\Mpch,150\Mpch]$ when evaluating the Gaussian convolution $\xi_0(s|\sigma)$ of \xilin{r} (equation~\ref{eq:gaussconv}) for various values of $\sigma$. We see that the truncated integrals achieve better than $1\%$ accuracy over the range $65\leq s/(\Mpch)\leq120$, suitable for model-agnostic BAO analyses.}
\label{fig:xilin-trunc-effects}
\end{figure}

Our final goal is related to BAO science, so we do not aim to reproduce the 2pcf over the full range of scales.  However, the range of scales which we must be able to model accurately are those which contribute to BAO analyses.  To a good approximation \cite{bharadwaj96, cs08}, the BAO observable is not \xilin{r} itself, but a smeared version of it that we denote $\xi_0(s|\sigma)$, where $s$ is the tracer separation in redshift space and $\sigma$ is a smearing scale that represents the impact of bulk velocity flows on the nonlinear 2pcf:
\begin{align}
\xi_0(s|\sigma) &= \int\frac{\der k}{k}\,\frac{k^3P_{\rm lin}(k;\bt)\,{\rm e}^{-k^2\sigma^2(\bt)/2}}{2\pi^2}\,\frac{\sin(ks)}{ks}
= \int\der\rr\,\Cal{N}(\ss-\rr;\sigma^2\mathbf{1})\,\xilin{r}\notag\\
&= \frac{1}{s}\int_0^\infty\frac{\der r\,r}{\sqrt{2\pi}\,\sigma}\left[\e{-(r-s)^2/2\sigma^2} - \e{-(r+s)^2/2\sigma^2}\right]\,\xilin{r}\,,
\label{eq:gaussconv}
\end{align}
where $\Cal{N}(\xx;\mathbf{\Sigma})$ is a Gaussian distribution in \xx\ with zero mean and covariance matrix $\mathbf{\Sigma}$, $\mathbf{1}$ is the 3-dimensional identity matrix, and $\sigma\sim10\Mpch$ or less. Therefore, for BAO analyses over the range $s/(\Mpch) = [65,120]$, we must have \xilin{r} accurate over the range $r/(\Mpch) = [30,150]$. This range was determined by requiring that, 
when the integral in \eqn{eq:gaussconv} is truncated to this narrower range in $r$ (rather than $0<r<\infty$), it returns a value that is within 1 percent of the exact (i.e. untruncated) value even when $\sigma = 12 \Mpch$ (see Fig.~\ref{fig:xilin-trunc-effects}). For some exploratory tests, we also use the reduced range $55\leq r/(\Mpch)\leq 125$, similar to the BAO range.

\subsection{Training and validation data}
\label{subsec:train-val-data}
We construct our training and validation data by treating combinations of values of $r$ and \bt\ chosen in the appropriate ranges as individual input samples $\xx=(r,\bt)$ of dimension $n_{\rm p}+1$, where $n_{\rm p}\leq7$ as mentioned earlier. The values of the cosmological parameters \bt\ are generated as an $n_{\rm p}$-dimensional Latin hypercube centered on the fiducial vector, with boundaries placed $\pm5\%$ away in each parameter direction (except for $w_0$ for which the boundaries are placed $\pm10\%$ away). For most of the parameters, except $\Omega_{\rm m}$ and $w_0$, this choice represents parameter ranges substantially larger than the $3\sigma$ confidence intervals inferred from the results of the Planck collaboration \cite{Planck18-VI-cosmoparam}. The role of $w_0$ is, in any case, primarily to change the multiplicative growth factor, which our analysis below is not sensitive to. The $\pm5\%$ range in $\Omega_{\rm m}$ roughly corresponds to the $2\sigma$ confidence interval reported by \cite{Planck18-VI-cosmoparam}. We will argue later that this, too, is sufficient for our purposes. Although there is no problem, in principle, with extending the parameter ranges beyond what we have chosen, in practice this would lead to heavier requirements on the sample sizes, NN sizes and consequently on computational resources. We have therefore not pursued such an extension in this work.

Below, we will primarily discuss examples in which only $\{\Omega_{\rm m},h\}$ are varied when constructing the basis \Cal{B}; for these, we independently generated $127$ pairs of $\bt=(\Omega_{\rm m},h)$ values for our training set and $22$ pairs for our validation set (this includes the fiducial \bt\ which is not included in the training set). In Appendix~\ref{app:smallrangetests}, we also consider an example in which all $7$ cosmological parameters were varied, for which we generated $300$ ($50$) independent vectors \bt\ for the training (validation) sets.

In principle, we could treat the separation $r$ similarly, and generate uniform random samples of $r$ in the appropriate range. We have found, however, that sampling $r$ on a finely spaced uniform grid leads to more stable results than with random sampling. We therefore define the $r$ samples using a linearly spaced grid in the range $[r_{\rm min},r_{\rm max}]$ with $\sim0.35\Mpch$ spacing, namely, with $351$ grid points for the default range $[30\Mpch,150\Mpch]$ and $201$ points for the shorter range $[55\Mpch,125\Mpch]$ mentioned above. For the default range, this leads to $44,577$ training and $7,722$ validation samples for the 2-parameter variation. For the shorter range, the corresponding numbers are $25,527$ training and $4,422$ validation samples. The full 7-parameter variation was only performed for the shorter range, with $60,300$ training and $10,050$ validation samples. In all cases, we have attempted to use data sets of minimal size and have checked that our results are not sensitive to increasing the sizes of any of the data.

These data sets are used for training individual networks as well as optimizing over various choices in network architecture and training, as described later in section~\ref{sec:neuralnetwork}. Note that, had we adopted the vectorized approach mentioned at the start of the section, sampling $r$ on the same values as chosen above to construct the 351-dimensional vector \rr\ would likely lead to much more complex NN architectures than we will deal with below, which would correspondingly be more difficult to train.

\subsection{Test data}
\label{subsec:test-data}
Having trained a NN, we wish to perform a sequence of three tests of increasing stringency to assess the absolute level of performance and generalization capacity of the NN, or in other words, the degree of completeness of the discovered basis. We describe these tests, which we label `elementary', `basic' and `stringent' in increasing order of stringency, later. Here, we describe the data sets that we will use for each test.

\begin{figure}
\centering
\includegraphics[width=0.5\textwidth]{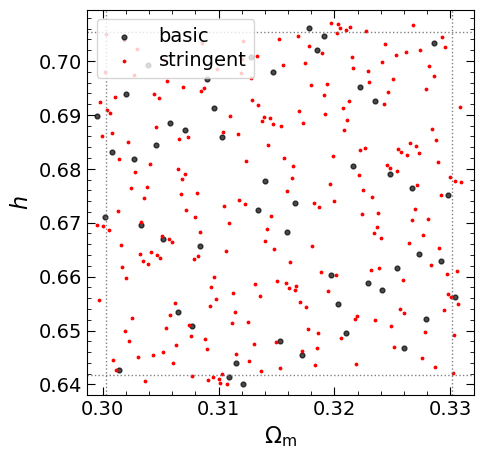}
\caption{2-dimensional (projections of the) \bt\ vectors used for the basic (black) and stringent (red) tests of our primary analysis, in the $(\Omega_{\rm m},h)$ plane. The horizontal and vertical dotted lines mark the boundary used for testing the edge accuracy of the model in section~\ref{sec:results}.}
\label{fig:test-data}
\end{figure}

The `elementary' test is based on the validation data already described above. For each of the two remaining tests, we consider $r$ sampled on a linearly spaced grid in the range $[r_{\rm min},r_{\rm max}]$ with $60$ points. For the `basic' test, we generate $n_{\rm basic}=50$ independent Latin hypercube samples of the $n_{\rm p}$-dimensional \bt\ vector, where $n_{\rm p}$ is the number of parameters varied when generating the training and validation data (so $n_{\rm p}=2$ for our primary analysis). We evaluate and store \xilin{r;\bt_a} for each $1\leq a\leq n_{\rm basic}$ on the $60$ linearly spaced $r$ values. Although the parameter vectors \bt\ comprising this `basic' data set are structurally and statistically identical to those in the validation data, we will see in section~\ref{subsec:tests} that they are used very differently. We create a similar data set for the `stringent' test, except that we now generate $n_{\rm stringent}=250$  independent samples of the \emph{full} $7$-dimensional \bt\ vector, regardless of how many parameters were varied in generating the training set. We emphasize that the specific \bt\ samples generated for the training, validation and two test data sets are all independent of each other. Fig.~\ref{fig:test-data} shows 2-dimensional projections of the \bt\ vectors used for the basic (black) and stringent (red) tests of our primary analysis.

\section{Neural network}
\label{sec:neuralnetwork}
\subsection{\biseq\ Architecture}
\label{subsec:arch}
A naive ML approach to this problem might be as follows:
\begin{itemize}
\item Train a fully-connected NN having $L$ hidden layers with the $(n_{\rm p}+1)$-dimensional input $(r,\bt)$ and scalar output \xilin{r;\bt}, with a squared loss $\Cal{L}=\sum(\hat\xi-\xi_{\rm true})^2$ using the ground truth $\xi_{\rm true}$ and its approximation $\hat\xi$.
\item The output of the last hidden layer (of width $n_\ast$, say) is now a collection of functions $\{a_i(r,\bt)\}_{i=1}^{n_\ast}$.
\item Average each of these functions over the parameters \bt\ to obtain a set of functions $\{A_i(r)\}_{i=1}^{n_\ast}$, where
\beq
A_i(r) = \avg{a_i(r;\bt)}_{\bt}\,,
\label{eq:avg-basis}
\eeq
and the average is over the parameter range used for defining the training set. The functions $\{A_i(r)\}$ could then be treated as the required basis set \Cal{B}.
\end{itemize}

An obvious problem with this approach is that the information carried by the parameters \bt\ is not cleanly separated from that contained in $r$. The averaging operation in \eqn{eq:avg-basis} will then almost certainly erase some information from the putative basis set and decrease its completeness or generalization capacity. We will later report the results of such an exercise which confirms this expectation. We therefore pursue an alternative NN architecture in which the separability expected from \eqn{eq:funcapprox} is built in from the start.

\begin{figure}[h!]
\centering
\includegraphics[width=\textwidth]{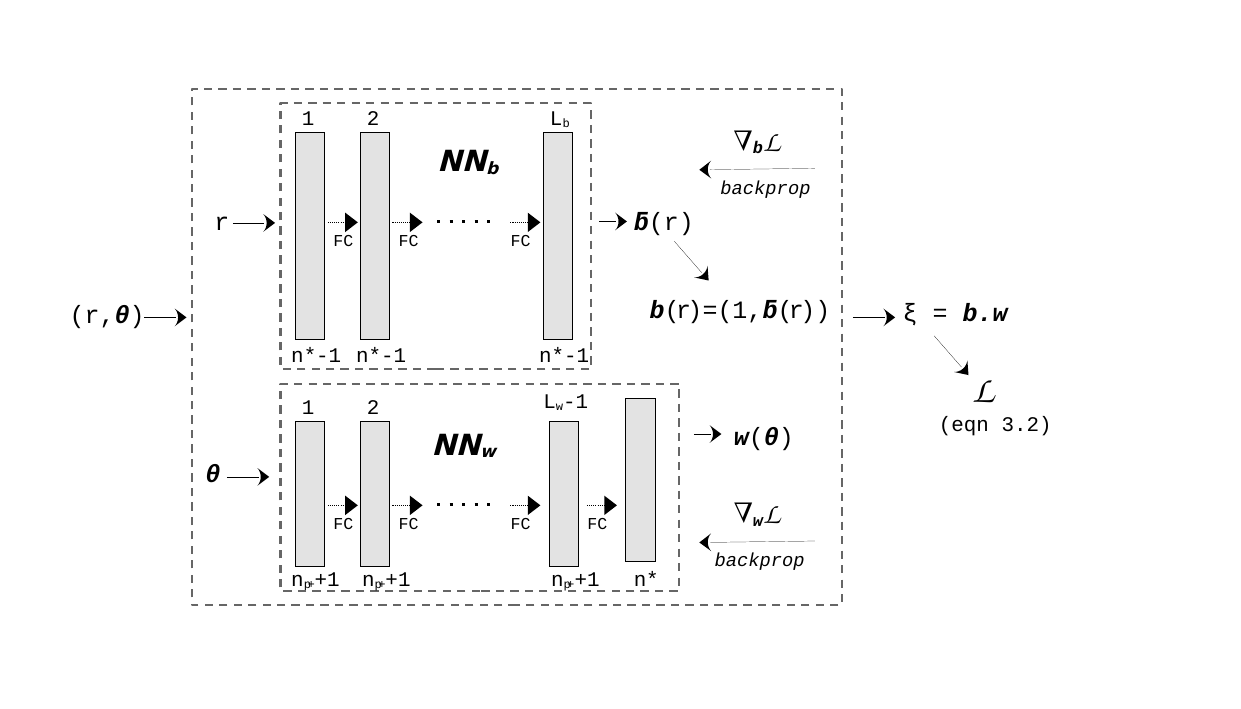}
\caption{Illustration of the \biseq\ architecture for approximating $\xilin{r;\bt}\approx\sum_{i=1}^{n_\ast}b_i(r)\,w_i(\bt)$. See text for details. The same setup generalizes to approximation problems of the type $f(\xx;\bt)\approx\sum_{i=1}^{n_\ast}B_i(\xx)\,W_i(\bt)$ for arbitrary vector input \xx.}
\label{fig:biseq-illustrate}
\end{figure}

The architecture we propose is shown in Fig.~\ref{fig:biseq-illustrate} and referred to as \biseq\ below. This essentially comprises of two fully-connected NNs working in tandem. The first network (\texttt{NNb}) takes a scalar value of $r$ as its input and produces an output $\tilde{\bb}(r)$ of dimension $n_\ast-1$. This is augmented with the constant unit function, so as to allow for a constant offset in the final result, leading to  an $n_\ast$-dimensional output $\bb(r)=(1,\tilde{\bb}(r))$. The second network (\texttt{NNw}) takes an $n_{\rm p}$-dimensional parameter vector \bt\ as its input and produces an $n_\ast$-dimensional output $\ww(\bt)$. The outputs of \texttt{NNb} and \texttt{NNw} are combined in the last layer of \biseq\ into the function approximation $\hat\xi(r;\bt)=\bb(r)\cdot\ww(\bt)$, which explicitly guarantees the required separability of the basis set $\bb(r)$ from the vector of coefficients $\ww(\bt)$. \texttt{NNb} and \texttt{NNw} are allowed to have different depths, $L_{\rm b}$ and $L_{\rm w}$, respectively. For simplicity, we use fixed widths for each layer of \texttt{NNb} (width $=n_\ast-1$) and each hidden layer of \texttt{NNw} (width = $n_{\rm p}+1$), with the last layer of \texttt{NNw} having width $=n_\ast$ by construction. We systematically search over a wide range of depths $L_{\rm b}$ and $L_{\rm w}$ as well as values of $n_\ast$ in the validation phase described later. We also vary the activation functions for the hidden layers among a few combinations of the hyperbolic tangent \texttt{Tanh} and the rectified linear unit \texttt{ReLU} \citep{nh10}. 

Treated as a whole, the input to the \biseq\ network is the $(n_{\rm p}+1)$-dimensional vector $(r,\bt)$, whose first component is fed into \texttt{NNb} and the remaining to \texttt{NNw}. Throughout, as is common practice, we train our networks on standardized outputs, where the standardization of the training samples of the `truth' is done by subtracting the mean and dividing by the standard deviation. Our code implementation automatically accounts for this standardization when producing the output of a trained \biseq\ instance. The code also provides convenience functions for extracting trained instances of \texttt{NNb} and \texttt{NNw} as standalone fully-connected NNs, giving easy access to, e.g., the basis set \Cal{B} which is the output $\bb(r)$ of \texttt{NNb}. We describe the training procedure for the \biseq\ network in the next section. Although we do not explore this here, it is also straightforward to extend the search range over individual architectures of \texttt{NNb} and \texttt{NNw} for generic function approximation problems of the type $f(\xx;\bt)\approx \bb(\xx)\cdot\ww(\bt)$ within our implementation of the \biseq\ framework.

\subsection{Training and validation}
\label{subsec:train-val-exc}
The \biseq\ loss function is defined as
\beq
\Cal{L} = \sum_s \left(\bb(r_s)\cdot\ww(\bt_s) - \xi_{\rm true}(r_s;\bt_s)\right)^2\,
\label{eq:biseq-loss}
\eeq
with $s$ labelling all the samples in the respective (mini-)batch. We train the \biseq\ network of a given architecture using mini-batch gradient descent (GD), with two simultaneously evaluated branches of back-propagation. At the output layer, the gradient $\nabla_{\bb}\Cal{L}$ with respect to \bb\ is back-propagated into the \texttt{NNb} branch, while $\nabla_{\ww}\Cal{L}$ goes into the \texttt{NNw} branch. Thereafter, each of these branches propagates the loss gradient independently as a standard fully-connected NN, with its own learning rate ($\ell_{\rm b}$ and $\ell_{\rm w}$, respectively, for \texttt{NNb} and \texttt{NNw}). 

We shuffle and split the training data into two chunks comprising $80\%$ and $20\%$, respectively, of the input training sample size $n_{\rm train}$. The larger of these is used for the GD updates using mini-batches of size $\sim\sqrt{0.8n_{\rm train}}$, while the loss $\Cal{L}_{\rm sub}(t)$ of the smaller sub-sample calculated at each epoch $t$ is used for deciding when to exit the GD loop. The exit checks start from epoch $t=1000$ onwards, and the GD loop is exited when the best-fit average slope $\avg{\p\Cal{L}_{\rm sub}/\p t}$ of the loss evolution, calculated using least squares on the last $1000$ epochs of this sub-sample, becomes positive. We have checked that the results are insensitive to minor variations in the duration over which the slope is calculated, but very small durations are not favoured due to increased noise, while very large durations lead to over-fitting. We use \texttt{adam} optimization \citep{kb14} for the back-propagation and GD updates, with coefficients $\beta_1=0.9$ and $\beta_2=0.999$ for the momentum calculation of the first and second moment, respectively, and $\epsilon=10^{-8}$ for the regularization of the second moment. We have checked that no other regularization such as weight decay or batch-normalization is required.

We use the validation data described in section~\ref{subsec:train-val-data} to select the appropriate \biseq\ network, as follows. We systematically vary $n_\ast$, the two depths $L_{\rm b}$ and $L_{\rm w}$ and the two learning rates $\ell_{\rm b}$ and $\ell_{\rm w}$ that define the network architecture, as well as the distribution of activation functions in each of \texttt{NNb} and \texttt{NNw}. For each architectural choice, we calculate the distribution $p(\delta)$ of relative residuals $\delta\equiv\hat\xi/\xi_{\rm true}-1$ between the ground truth $\xi_{\rm true}$ and its approximation $\hat\xi$, using the entire pool of samples in the validation data. The width\footnote{Not to be confused with the smearing scale in \eqn{eq:gaussconv}.} $\sigma$ of $p(\delta)$ is estimated as half the difference between the $84^{\rm th}$ and $16^{\rm th}$ percentiles of the distribution. We repeat this exercise $5$ times by changing the random number seed used for the mini-batch GD (i.e., changing the random order in which the algorithm is exposed to the training data). We choose the architecture which consistently yields $\sigma\leq0.05$ for at least 4 out of 5 repetitions, along with at least one instance having $\sigma < 0.01$ (i.e., sub-percent validation accuracy at $68\%$ confidence). For this best architecture, we repeat the training process $15$ times by changing the random number seed and store the network that gives the smallest value of $\sigma$. 

We have found, in general, that a \texttt{Tanh} activation works well for \texttt{NNw}, while a sequence of \texttt{ReLU} activations followed by \texttt{Tanh}, in a $3:1$ ratio, works well for \texttt{NNb}. The split between \texttt{ReLU} and \texttt{Tanh} for \texttt{NNb}, in particular, ensures sufficient generalizability while producing smooth outputs for $\bb(r)$.
When using $\bt=(\Omega_{\rm m},h)$, so that $n_{\rm p}=2$, the best architecture resulting from the training+validation phase and having the smallest basis size has $n_\ast=9$ (i.e., $8$ non-trivial basis functions), $L_{\rm b}=20$ and $L_{\rm w}=6$. Here, \texttt{NNb} comprises $16$ layers with \texttt{ReLU} activation, followed by $4$ layers with \texttt{Tanh} activation, each of width $8$, while \texttt{NNw} comprises $5$ layers of width $3$ followed by a last layer of width $9$, each with \texttt{Tanh} activation (see also the discussion of the discovered basis functions in section~\ref{sec:results}). The training is achieved with $\ell_{\rm b}=3\times10^{-6}$ and $\ell_{\rm w}=10^{-5}$, with $1477$ NN weights optimized.

In addition to these `deep+narrow' architectures, we also explored `shallow+wide' architectures having $L_{\rm b},L_{\rm w}=2$ or $3$ but having large widths $\gtrsim 100$. In general, we have found that the `deep+narrow' architectures generalize much better for our problem, in addition to being substantially easier to train due to the exponentially smaller number of optimized NN weights.

\subsection{Tests}
\label{subsec:tests}
The generalization capacity of the chosen trained NN (or the completeness of  the discovered basis set) can be assessed using the `elementary', `basic' and `stringent' tests alluded to earlier, which we now describe. In each of these tests, we again calculate the distribution of relative residuals $\delta\equiv\hat\xi/\xi_{\rm true}-1$ between the ground truth and its approximation, now evaluated on the various testing data. 

In the `elementary' test, we compute the residuals on the validation samples described in section~\ref{subsec:train-val-data}. This is, of course, not necessarily a good indicator of the NN's generalization capacity, since the validation data are part of the process of identifying the best amongst the trained NNs. We therefore perform the `basic' and `stringent' tests, as follows. For each of these, we consider the $n$ \emph{functions} \xilin{r;\bt_a}, $1\leq a\leq n$, where $n=n_{\rm basic}$ ($n=n_{\rm stringent}$) for the `basic' (`stringent') test. These $n$ functions are individually approximated as linear combinations of the basis functions extracted from the optimal NN at the same $60$ values of $r$, using a standard least squares approach (i.e., assuming a constant but unspecified error on \xilin{r;\bt_a}). We then calculate the residuals $\delta$ between each ground truth $\xi_{\rm true}=\xilin{r;\bt_a}$ and its approximation $\hat\xi = \sum_{i=1}^{n_\ast}\,w^{(a)}_i\,b_i(r)$, where $\{w^{(a)}_i\}_{i=1}^{n_\ast}$ are the corresponding best fitting basis coefficients. These residuals are pooled together for all $r$ and $\bt_a$ values, and the resulting distribution is recorded. 

Some comments are in order. Firstly, an important difference between the `elementary' and `basic' tests is that the former uses coefficients produced by the trained \texttt{NNw} instance, while the latter ignores \texttt{NNw} altogether and estimates coefficients using a direct least squares optimization. The `basic' and `stringent' tests share this property of deriving coefficients from least squares comparisons with the target output or `truth'. Secondly, from the descriptions of the `basic' and `stringent' test data sets in section~\ref{subsec:train-val-data}, it should be clear that the `basic' test assesses the generalizability of the discovered basis to parameter combinations \bt\ different from the specific realisations used in the training and validation data, while the `stringent' test assesses the overall completeness of the basis when \emph{all} parameters are allowed to vary. Finally, for the `basic' test applied to the result of the \biseq\ architecture, the best fit coefficients $\{w_i^{(a)}\}_{i=1}^{n_\ast}$ for each $\bt_a$ should, in principle, agree with the coefficients $\ww(\bt_a)$ produced as the output of the \texttt{NNw} network. This might not be the case in practice, however, since the mini-batch GD algorithm is not guaranteed to reach the true optimum of the loss function to machine precision, due to its dependence on the training data size, the order in which samples are shown to it, the details of the exit criterion, etc. In any case, for the `stringent' test applied to the same result, there is no unique mapping between the fitted coefficients and the output of \texttt{NNw}, since the latter takes a \emph{lower}-dimensional vector as its input. Regardless of the test, the values of the coefficients \ww\ are not particularly relevant to us in the present work, since our goal is only to determine whether the \emph{basis functions} are sufficiently generalizable/complete. The inverse problem, that of \emph{inferring} values of \bt\ in some cosmological model, given some constraints on \ww, is definitely interesting and we will pursue this in the near future.


\section{Results}
\label{sec:results}
We have performed the training, validation and testing exercise described in the previous section for a number of configurations of the range $[r_{\rm min},r_{\rm max}]$ and choice of input parameters \bt. 
In principle, our training data should allow all 7 parameters discussed in section~\ref{subsec:cosmo} to vary. The resulting networks can be somewhat cumbersome and difficult to train, however (we discuss an example in Appendix~\ref{app:smallrangetests}). Moreover, on physical grounds we also expect the combination $\{\Omega_{\rm m},h\}$ to be the most relevant in deciding the shape of the 2pcf near the BAO feature \citep[cf., e.g., the shape parameter $\Gamma=\Omega_{\rm m}h$ discussed by][]{bbks86}. A training set based on varying only $\bt=(\Omega_{\rm m},h)$ should therefore be sufficient to produce a NN that can generalize to varying \emph{all} 7 parameters. While we confirm this expectation below, one might also expect $\Omega_{\rm b}$ to be an equally important parameter since it affects the height of the BAO peak relative to the broadband shape. We return to this point below.

\begin{figure*}[h!]
\centering
\includegraphics[width=0.43\textwidth]{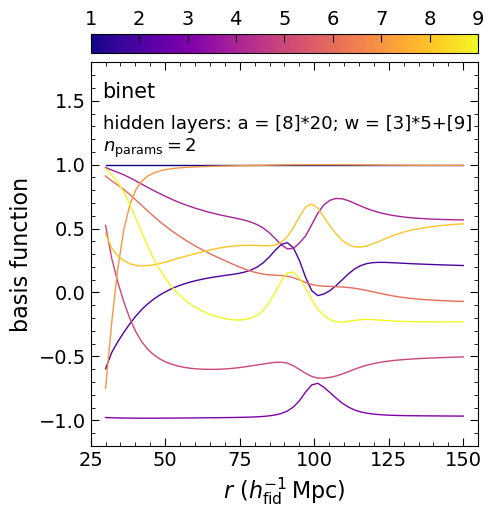}
\includegraphics[width=0.44\textwidth]{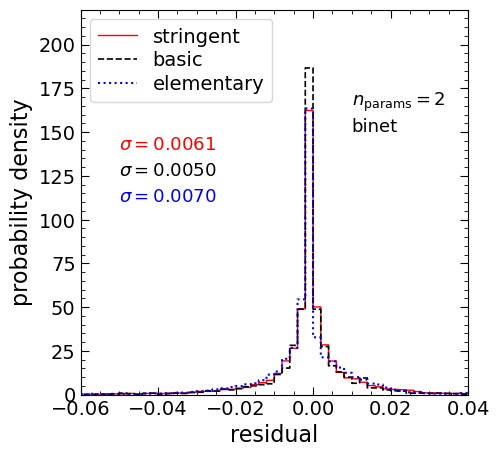}
\caption{\emph{(Left panel):} Basis functions discovered by applying the training and validation procedure described in section~\ref{subsec:train-val-exc}, where the training data were obtained by varying the 2 parameters $\{\Omega_{\rm m},h\}$, with \xilin{r} evaluated on the full range $30\leq r/(\Mpch) \leq 150$, as described in section~\ref{subsec:train-val-data}. The colour bar indicates integer labels assigned to each function. \emph{(Right panel):} Distributions of residuals obtained for the `elementary' (blue dotted), `basic' (black dashed) and `stringent' (red solid) tests described in section~\ref{subsec:tests}, using the corresponding data sets described in section~\ref{subsec:test-data}. See section~\ref{sec:results} for a discussion of the results.}
\label{fig:biNN2p-basis-residuals}
\end{figure*}

The \emph{left panel} of Fig.~\ref{fig:biNN2p-basis-residuals} shows the $9$ basis functions \Cal{B} resulting from the training and validation exercise described in section~\ref{subsec:train-val-exc} when varying $\bt=(\Omega_{\rm m},h)$. These are restricted to the range $[-1,1]$ due to our choice of a \texttt{Tanh} activation in the last layer of \texttt{NNb}. Using a \texttt{ReLU} activation at the last layer would instead lead to unbounded functions. While this is not a problem \emph{per se}, we find it easier to deal with bounded functions in the basis set \Cal{B}, transferring the burden of setting the overall amplitude of the predicted output to the weights \Cal{W}.\footnote{By this reasoning, in principle, the weights output by \texttt{NNw} should be allowed to be unbounded, e.g. by using a linear or \texttt{ReLU} activation instead of our choice of \texttt{Tanh} in its last layer. In practice, however, the standardization mentioned in section~\ref{subsec:arch} means that a \texttt{Tanh} last activation works well for both \texttt{NNb} and \texttt{NNw}. When performing least squares fits for the weights instead, as done in the `basic' and `stringent' tests for example, the fitting procedure itself accounts for the overall amplitude of the weights.} The shapes of the basis functions are inherently interesting too; we discuss these in Appendix \ref{app:basisfuncs}.

The \emph{right panel} of Fig.~\ref{fig:biNN2p-basis-residuals} shows the distribution of residuals $p(\delta)$ for the `elementary', `basic' and `stringent' tests described in \ref{subsec:tests}. By construction, the width $\sigma$ of $p(\delta)$ is sub-percent for the `elementary' test. The width is also sub-percent for the `basic' test, and smaller than that for the `elementary' test. This is already encouraging and indicates that the discovered basis generalizes nicely beyond the specific realization of the training and validation data sets. Most interestingly, though, the `stringent' test \emph{also} has a sub-percent width $\sigma\sim0.6\%$, comparable to the other two tests. Since \emph{all} 7 parameters are varied in generating the data set for the `stringent' test, this clearly demonstrates that the discovered basis is sufficiently complete to cover the entire target range and dimensionality of parameter variation. For comparison, we have checked that fitting the stringent test data to polynomials of degree 8 -- which have the same number of (now, monomial) basis functions as \Cal{B} -- gives a corresponding width of residuals $\sigma\sim14.5\%$, a factor $24$ worse than for \Cal{B}. Increasing the polynomial degree to 17 (i.e., allowing \emph{twice} as many functions as in \Cal{B}) improves the quality of the residuals, but still gives a width of $\sim2.2\%$, a factor $\gtrsim3$ worse than for \Cal{B}. The monomial basis is evidently far inferior to the \Cal{B} discovered in our approach.

We mentioned above that, \emph{a priori}, a variation in $\Omega_{\rm b}$ might be expected to be as relevant in producing sufficiently varied shapes of the BAO feature as are $\{\Omega_{\rm m},h\}$. Additionally, the slope $n_{\rm s}$ of the initial power spectrum might also be expected to play an independent role. Indeed, several emulation analyses focus on the combination $\{\omega_{\rm b},\omega_{\rm c},n_{\rm s}\}$ as the primary set of variables, where $\omega_i=\Omega_ih^2$ and  $\omega_{\rm c}=\omega_{\rm m}-\omega_{\rm b}$ \citep{eggemeier+23,bcv24}. The results of the `stringent' test explicitly show that, by varying $\{\Omega_{\rm m},h\}$ alone, our network learns sufficient information about the nature of variation of the 2pcf shape so as to describe the effect of \emph{all} relevant parameters. The fact that we did \emph{not} train on all these parameters, means that our `stringent' test is really stringent. Also, since our parameter variations are mostly $\pm5\%$ by choice, it is also clear that varying $\Omega_{\rm m}\,(h)$ in the $\sim2\sigma\,(\sim6\sigma)$ confidence interval of the Planck constraints \cite{Planck18-VI-cosmoparam} is sufficient to describe the effect of varying other parameters in confidence intervals exceeding $6\sigma$ for parameters such as $\{n_{\rm s},\Omega_{\rm b}\}$. Parameters such as $\{A_{\rm s},w_0\}$ are less worrisome due to their primarily multiplicative effect on \xilin{r}. Below, we address the concern that a $2\sigma$ range for exploring the effect of $\Omega_{\rm m}$ itself might not be wide enough.

\begin{figure*}[h!]
\centering
\includegraphics[width=0.45\textwidth]{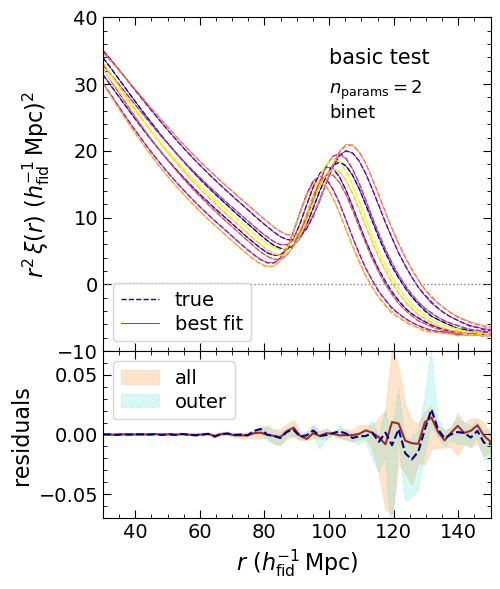}
\includegraphics[width=0.45\textwidth]{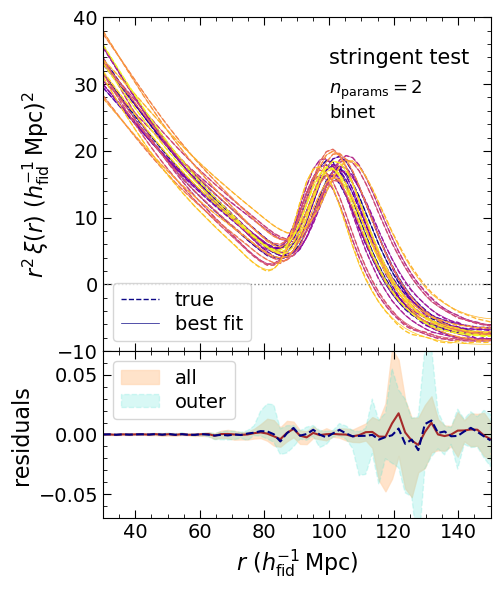}
\caption{\emph{(Upper panels):} Comparison of the best-fit (solid) and true (dashed) functions \xilin{r;(\Omega_{\rm m},h)} obtained in the `basic' \emph{(left panel)} and `stringent' tests \emph{(right panel)}, using the basis functions shown in Fig.~\ref{fig:biNN2p-basis-residuals}. We display results for 10 (25) randomly chosen parameter combinations out of 50 (250) in the \emph{left (right)} panels. The horizontal dotted line in each panel marks the value zero. \emph{(Lower panels):} Median (thick lines) and central 68\% interval (bands) of the distribution of residuals $\delta$ calculated at each scale $r$. The solid brown line and brown band for each test shows the residual distribution when using all the \bt\ vectors (i.e., 50 and 250, respectively, for the basic and stringent tests). The dashed blue line and blue band shows residuals when restricting to \bt\ vectors in the boundary of the 2-dimensional projection shown in Fig.~\ref{fig:test-data}, which comprise $\sim11\%$ of the respective samples.}
\label{fig:biNN2p-tests}
\end{figure*}

Our definition of $p(\delta)$ and its width $\sigma$ as a diagnostic for the performance of the discovered basis is agnostic to the specific values of $r$ where the residual $\delta$ is measured. Since one is typically interested in accuracy at specific choices of $r$, such as the peak, linear point, zero-crossing etc., it is worth examining the actual shapes of the approximations in detail.
The \emph{upper panels} of Fig.~\ref{fig:biNN2p-tests} show some example comparisons between the ground truth (dotted curves) and the least-squares best-fit approximation using the discovered basis (solid curves), where the ground truth is randomly selected from the data generated for the `basic' (\emph{left panel}) and `stringent' test (\emph{right panel}). Visually, these comparisons show excellent agreement, with noticeable deviations typically occurring at large separations $r\gtrsim120\Mpch$. 
The \emph{lower panels} show the median (continuous curves) and central $68\%$ interval (shaded bands) for the corresponding residuals $\delta$ for each test data set, calculated separately at each $r$. The residual distributions are shown separately for the entire test data set (brown) and when restricting to \bt\ vectors lying in the boundary region shown in Fig.~\ref{fig:test-data} (blue). We see that, in general, the residuals remain sub-percent at small $r$, increase to $\gtrsim5\%$ in the vicinity of the zero-crossing (which is entirely expected, since small variations in the location of the zero-crossing can lead to near divisions by zero in calculating relative residuals), and saturate at $\sim2\%$ at $r\gtrsim140\Mpch$. Importantly, though, the behaviour of the residuals for vectors in the edge of the test data are very similar to those for the full data set, for both `basic' and `stringent' test data. This suggests that the basis functions discovered by our analysis are relatively robust to small variations in the training parameter ranges. We note that this edge test is not as strict as it could have been for the `stringent' data set, since we defined the `edge' of the data only in a 2-dimensional projection in Fig.~\ref{fig:test-data}. The test nevertheless builds confidence that, not only are our choices reasonable, but that the discovered basis might even generalize with reasonable accuracy to \bt\ values \emph{outside} our chosen training range. We leave a fuller exploration of this issue to future work.

\begin{figure*}[h!]
\centering
\includegraphics[width=0.45\textwidth]{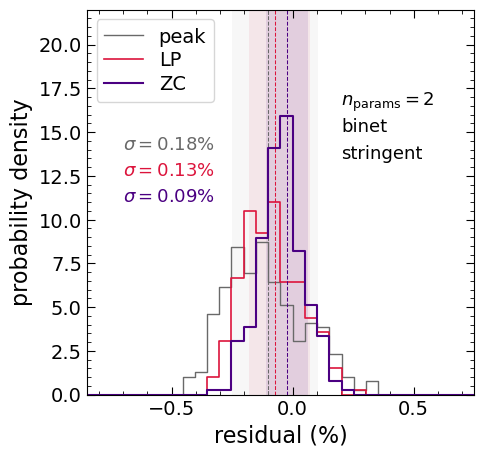}
\caption{Percentage residuals for recovery of peak (gray), linear point (LP; red) and zero-crossing (ZC; blue) of \xilin{r}, using data generated for the `stringent' test. Histograms show the probability density functions, vertical dotted lines indicate the medians and vertical shaded bands mark the central $68\%$ confidence intervals of the respective residuals. Labels indicate the widths of each distribution.}
\label{fig:biNN2p-scaleresids}
\end{figure*}

To quantify the scale-dependence of the residuals further, Fig.~\ref{fig:biNN2p-scaleresids} shows the performance of the discovered basis in predicting the location of the peak, linear point (LP) and zero-crossing (ZC) of \xilin{r} using the `stringent' test data. Each histogram shows the distribution of the percentage residual $\delta_{s}\equiv 100\times(\hat s/s_{\rm true}-1)$ between the ground truth $s_{\rm true}$ and its approximation $\hat s$, with $s$ labeling each of the three scales mentioned above. We see excellent recovery of each of the three scales. The LP residuals have a median and central $68\%$ confidence interval of $\delta_{\rm LP}=-0.07^{\,+0.14}_{\,-0.11}\,\%$, about a factor $3$ smaller than the expected LP errors from a DESI LRG-like survey \citep{LP2016,ps23}. The peak residuals are about a factor $1.4$ larger in magnitude than the LP residuals, with $\delta_{\rm peak}=-0.10^{\,+0.21}_{\,-0.15}\,\%$; this is consistent with earlier results indicating that the LP can be estimated more accurately than the peak \citep{LP2016,lnps24}. The ZC residuals, on the other hand, perform better than the LP, with $\delta_{\rm ZC}=-0.023^{\,+0.086}_{\,-0.089}\,\%$. This is particularly interesting for upcoming analyses of DESI data \citep{Prada2011, bkBAO,lnps24}. 

For completeness, we note that, to calculate the peak and LP, we used an iterative search with an adaptive fine grid (largest grid size $\Delta r_{\rm max}=0.02\Mpch$), having an estimation accuracy of better than $\sim0.025\%$. We can do this because our NN architecture allows us to evaluate the basis functions at arbitrary $r$ within the training range. For the ZC, we instead applied a simple numerical search for the zero crossing, refined with one linear interpolation, on the same 60 $r$ values used in the test data sets, giving an accuracy (using Rolle's theorem) of better than $\sim\frac18(\Delta r)^2\der^2\xilin{r}/\der r^2 \sim \frac18(2)^2\times 8\times10^{-6}\sim0.0004\%$, where we set $\Delta r=2\Mpch$ for the test data and numerically estimated a typical value for the second derivative of the 2pcf in the fiducial cosmology in the range $110<r/(\Mpch)<135$ where the ZC is expected to lie. These estimation accuracies are much better than the typical NN errors reported above, ensuring that our results are numerically robust.

We study the impact of the choice of parameters varied during training and validation in Appendix~\ref{app:smallrangetests}. For completeness, we have also explored (but not displayed results for) the naive approach mentioned at the start of section~\ref{subsec:arch}, with an `averaged' basis set obtained as in \eqn{eq:avg-basis}. In this case, although it was possible to obtain $\lesssim1\%$ validation accuracy, the `basic' and `stringent' tests on the resulting basis yielded accuracies of $\gtrsim3\%$ for $\bt=(\Omega_{\rm m},h)$, i.e., about a factor $5$ worse than the \biseq\ results reported above. This is consistent with the expected loss of information in this naive approach as discussed in section~\ref{subsec:arch}.
All these results support our use of the \biseq\ architecture along with the choice of $\bt=(\Omega_{\rm m},h)$ as being a minimal parameter set for obtaining a sufficiently complete basis \Cal{B} for approximating the linear 2pcf on the required range of separations.

\section{Discussion}
\label{sec:discuss}

\subsection{Uniqueness and orthogonality}
\label{subsec:unique}
The basis \Cal{B} we have discovered in Fig.~\ref{fig:biNN2p-basis-residuals} is not unique, since the specific functions $\bb(r)$ returned by \texttt{NNb} for a particular network architecture constitute one among several acceptable solutions explored by the training algorithm. This is not particularly problematic, since we are ultimately interested in \emph{some} example of a basis set which is sufficiently generalizable. Also, 
since any set of linearly-independent linear combinations of a basis is an equally acceptable basis, we can always tolerate some level of mixing of the basis functions. 

Another interesting observation relates to the orthogonality of the basis functions. In the absence of any notion of a metric provided by, e.g., a data covariance matrix (see below) or other weights such as those appearing in Sturm-Liouville theory, a natural definition of a norm between functions $f(r)$ and $g(r)$ might be
\beq
\avg{f\,|\,g} \equiv \int_{r_{\rm min}}^{r_{\rm max}}\der r\,r^2\,f(r)\,g(r)\,,
\label{eq:func-norm}
\eeq
where the $r^2$ factor appears when projecting pair counts in 3 dimensions onto a chosen basis.  The discovered basis in Fig.~\ref{fig:biNN2p-basis-residuals} is then visibly non-orthogonal, i.e., it is not true that $\avg{b_i\,|\,b_j}=0$ for $i\neq j$. E.g., there are some purely positive and other purely negative functions, which therefore cannot be orthogonal to each other or to the constant function. By itself, this is not a major issue since, e.g., the monomial basis routinely used in BAO studies is also not orthogonal. If needed, one can always perform the equivalent of the Gram-Schmidt procedure on \Cal{B} to obtain an equivalent orthonormal basis. As we discuss below, however, orthogonality becomes more relevant in the context of a specific model and/or data covariance structure. To the extent that we are interested in a purely theoretical, model-agnostic basis set \Cal{B}, such an orthogonalization is an unnecessary detail at this stage.

\subsection{Connection to related approaches in compression}
\label{subsec:compare_lit}

The identification of a basis set \Cal{B} comprising a finite, small number of functions makes our approach an exercise in \emph{functional} compression. Namely, by discovering and fixing the basis, we have replaced the function \xilin{r} with a corresponding small set of numbers $\Cal{W}=\{w_i\}$ that lead to the approximation $\xilin{r}\approx\sum_iw_i\,b_i(r)$. It is interesting to compare our framework with related \emph{data} compression approaches from the literature. To put things in context, it is helpful to first think of a typical BAO-based inference exercise as the sequence
\beq
\boxed{\textrm{model},\bt_{\rm model}} \longrightarrow \boxed{\xilin{r},\bt_{\rm agnostic}} \longrightarrow \boxed{\textrm{inference}} \longleftarrow\, \boxed{\textrm{data}}\,,
\label{seq:typical}
\eeq
where the first arrow indicates the calculation of \xilin{r} in a typical cosmological model with parameters $\bt_{\rm model}$ and the second arrow indicates all the subsequent steps (nonlinearities, redshift space effects, likelihood evaluation, etc.) involved in the inference of $\bt_{\rm model}$, which of course also involves the data as indicated by the last arrow. The second step may also involve some additional parameters which we denote $\bt_{\rm agnostic}$ since these will typically not depend on the specific details of the model that defines \xilin{r}. E.g., $\bt_{\rm agnostic}$ may include the average linear bias $b$ of the tracer population being modelled.

In this language, our intended use of the basis \Cal{B} would be the sequence
\beq
\boxed{\Cal{B},\Cal{W}} \longrightarrow \boxed{$\xilin{r}$,\bt_{\rm agnostic}} \longrightarrow \boxed{\textrm{inference}} \longleftarrow \,\boxed{\textrm{data}}\,,
\label{seq:agnostic}
\eeq
where the first step now uses the fixed basis \Cal{B} with variable weights \Cal{W} to predict \xilin{r}, and the second step augments this with the collection of model-agnostic parameters $\bt_{\rm agnostic}$ to perform the inference step. In this case, in addition to parameters like the tracer bias $b$, the collection $\bt_{\rm agnostic}$ can also include parameters that would, in principle, have been predicted by a specific model such as $\Lambda$CDM in the model-dependent sequence \eqref{seq:typical}, but are left free in the model-agnostic approach \eqref{seq:agnostic}.
E.g., following \cite{ps23}, these could be $\{\sigma_{\rm v},f\}$, where $\sigma_{\rm v}$ is the velocity dispersion in linear theory and $f$ is the linear growth rate. 

We will compare our framework with two classes of approaches from the literature: (i) those based on principal components analysis (PCA) of the data covariance \citep[e.g.,][]{connolly+95,Lahav2009,lnps24} and (ii) model-dependent data compression schemes such as \textsc{moped} \citep{moped-hjl00} and its generalizations \citep{aw18} as well as ML-based successors such as information maximizing neural networks \citep{imnn-clw18} and related frameworks \citep{makinen+24}.

\subsubsection{Connection to PCA}
In the PCA approach, a sampling of $r$ as a vector \rr\ (or the redshift space \ss) is pre-decided, and the eigenvectors of the corresponding covariance matrix of, say, the monopole $\xi^{(0)}$ of the redshift-space 2pcf measured on \rr\ or \ss\ are a natural choice for the basis \Cal{B} in fitting the shape of the linear 2pcf \citep{lnps24}. Although model-agnostic like our approach, the PCA method relies on knowledge of the target data set, with the resulting basis being strictly valid only for a particular choice of survey definition, tracer selection, etc. Typically, one retains only a subset of the eigenbasis in any fitting exercise, so as to focus on relatively smooth functions that account for most of the intrinsic variation in the shape of the 2pcf, while ignoring the degrees of freedom associated with short-wavelength noise.

Our approach, on the other hand, constructs a purely theoretical basis \Cal{B} for \xilin{r}, and can consequently be applied to a variety of data sets, such as different surveys and/or combinations of the 2pcf multipoles $\xi^{(\ell)}$ for $\ell=0,2,4$, as outlined by \cite{ps23}. For specific choices of such data sets, a subset of the basis functions $b_i(r)$ may be relevant for the modelling, e.g., those selected using Bayesian evidence-based criteria as discussed by \cite{ps22}. When data errors are sufficiently small, one may even imagine all $9$ basis functions from Fig.~\ref{fig:biNN2p-basis-residuals} being relevant.

In the language introduced above, both PCA and our approach correspond to the sequence \eqref{seq:agnostic}. While our approach constructs the basis using the ML-based technique described earlier, without relying on any data set, the PCA approach explicitly uses the data covariance to construct the basis. Further, being purely theoretical, our discovered basis directly applies to real-space separations $r$ relevant for the linear 2pcf \xilin{r}, while PCA basis sets constructed using, say, $\xi^{(0)}(s)$ might need further processing in order to be applicable to \xilin{r}.

\subsubsection{Connection to least squares estimation of 2pcf from data}
Recent work has discussed the utility of non-trivial basis functions in the \emph{estimation} of the 2pcf in a data set \cite{tessore18,s-fh21}. The idea here is to first think of the usual observed 2pcf estimate in bins of separation $s$ as a least squares \emph{projection} of tracer pairs onto top-hat functions in $s$ (followed by a linear combination of these projections), and then generalize from top-hat functions to an arbitrary set of functions. Among several advantages of such an approach is the opportunity to describe the \emph{observed} 2pcf using only a limited number (say, a dozen) of measured basis coefficients, as opposed to the typically larger number (few dozens) of measured binned counts one usually works with. The resulting decrease in data dimensionality arising from this compression has obvious benefits when estimating the covariance matrix of the data vector. The basis \Cal{B} discovered above, although it arises in a purely theoretical context, should also be directly applicable to such an exercise in data compression. We intend to pursue this in the near future.

\subsubsection{Connection to MOPED and successors}
A widely used class of compression schemes involves either linear or non-linear compression of a given data set, in the context of a specific model. These are generic schemes that are also, in principle, applicable to the BAO problem of our interest. The first of these was the linear compression scheme \textsc{moped} \citep{moped-hjl00} (see also \cite{tth97}), in which one constructs a basis set of dimensionality equal to the number $M$ of parameters by finding $M$ linear combinations of the data, each of which maximizes the Fisher information for a specific parameter. Under certain conditions,\footnote{As derived by \cite{moped-hjl00}, for the compression to be lossless, the data covariance matrix must be independent of the parameters and the likelihood defined by the full data set should be close to a Gaussian in the vicinity of its peak, which itself should be close to the chosen fiducial model.} this information remains the same as that contained in the original data set, making the compression explicitly lossless. In a later study, \citep{aw18} generalized this concept by widening the class of situations that allow (nearly) lossless compression. Simultaneously, recent advances in ML have led to further generalizations of this idea to \emph{neural} compression schemes, where a NN is trained to construct \emph{nonlinear} summaries of the data that maximize the Fisher information on the model parameters; these are referred to as information maximizing neural networks or IMNNs \citep{imnn-clw18}.

All of these approaches share a common theme, namely, they work for a specific model having, say $M$ parameters and a specific data set of dimensionality, say $N$, where typically $N\gg M$. Each scheme then aims to compress the data to a size $\gtrsim M$ (ideally, $=M$), while being as lossless as possible. In this sense, these compression schemes work within the sequence \eqref{seq:typical} above, requiring both model and data to be specified. This makes it clear that, although our model-agnostic approach to discover \Cal{B} can be thought of as an exercise in neural compression, it is conceptually very different from the compression schemes mentioned here. Firstly, our approach is not tied to any particular data set. Secondly, although the explicit construction of \Cal{B} requires a model, we saw that varying only a couple of parameters to construct the training data leads to a highly complete basis that generalizes to variations of many other parameters. In other words, exploring a few models for \xilin{r} is sufficient to discover a basis relevant for a much wider class of models. Overall, this means that it is not straightforward to compare our approach with \textsc{moped} or its successors as a generic compression scheme. However, it would still be interesting to restrict attention to, say, 2pcf monopole measurements of a specific data set such as the DESI LRG sample and compare the basis \Cal{B} of Fig.~\ref{fig:biNN2p-basis-residuals} with the results of the \textsc{moped} or IMNN approach for increasingly complex $\Lambda$CDM models. Although outside the scope of the current work, we will report the results of such an exercise in a future publication.

Finally, as we pointed out earlier, our approach produces a basis \Cal{B} that is not orthogonal with respect to any simple metric. A nice feature of \textsc{moped} and related approaches is that their resulting basis sets are orthogonal with respect to the data covariance matrix $C$ used as a metric: $\sum_{\alpha,\beta}b_m(r_\alpha)\,C_{\alpha\beta}\,b_n(r_\beta)=\delta_{mn}$. This orthogonality ensures, e.g., that the corresponding compressed scalars are uncorrelated, which significantly simplifies the compressed likelihood, especially when the original likelihood is locally Gaussian near the fiducial model. In our case, when using \Cal{B} in the context of a particular data set with covariance matrix $C$, it should be possible to perform the Gram-Schmidt orthogonalization of \Cal{B} with the metric $C$ \citep[e.g., equation~14 of][replacing their model derivatives $\mu_{,\alpha}$ with our non-orthogonal basis functions]{moped-hjl00} after suitably mapping the data space variable (such as the redshift space separation $s$) to the real space $r$ relevant for the linear 2pcf. It will be interesting to ask how significant the impact of this orthogonalization is when using \Cal{B} to forward model, say, the full set of nonlinear 2pcf multipoles $\xi^{(\ell)}(s)$, an exercise we also leave to future work.

\subsection{Beyond $\Lambda$CDM}
\label{subsec:beyondLCDM}
The fact that a training data set constructed by varying only two parameters $\{\Omega_{\rm m},h\}$ around our fiducial flat $\Lambda$CDM model leads to a basis \Cal{B} that successfully describes the variations of all other curved $w$CDM parameters explicitly demonstrates that it is possible to discover a common basis for a wide class of models. In our case, one can think of the training set as having explored a small subset of flat $\Lambda$CDM models, while the discovered basis applies to not only the full flat $\Lambda$CDM set, but also to curved $\Lambda$CDM, flat $w$CDM, and curved $w$CDM. Although these are nested models, this success leads us to speculate that the \emph{same} basis \Cal{B} from Fig.~\ref{fig:biNN2p-basis-residuals} might equally well describe certain classes of modified gravity models sufficiently close to our fiducial $\Lambda$CDM model, namely those in which the BAO feature in linear theory is at least qualitatively similar to that in $\Lambda$CDM (a single peak followed by a zero-crossing). Similarly, since the BAO feature is sensitive to neutrino mass \cite[see, e.g., fig.~1 of][]{parimbelli+21}, the basis \Cal{B} might provide good descriptions of massive neutrino models too.

It will be interesting to test how far this expectation is satisfied in realistic models of modified gravity and/or massive neutrinos. For the latter, this can be easily checked using \textsc{class}, and it should also be straightforward for modified gravity models using suitable Boltzmann codes such as \textsc{hi\_class} \citep{hiclass-I-zuma+17,hiclass-II-bsz20}, an exercise we will take up in the near future. If successful, this would dramatically widen the class of models covered by \Cal{B}.

\section{Conclusion}
\label{sec:conclude}
We have developed a machine learning (ML) approach which exploits the fact that, although neural networks (NNs) considered for function approximation use highly nonlinear transformations of the input data, their final output is typically a linear combination of variables.  Therefore, one can think of the output of the penultimate layer of the NN as providing a basis set.  With this in mind, we approximate the linear 2pcf \xilin{r;\bt} using a linear combination (equation~\ref{eq:funcapprox}) of $n_\ast$ model-independent basis functions $\Cal{B}=\{b_i(r)\}_{i=1}^{n_\ast}$ with model-dependent weights $\Cal{W}=\{w_i(\bt)\}_{i=1}^{n_\ast}$, where both $\cal{B}$ and $\cal{W}$ were provided by the NN. 

Our initial goal was to discover a set \Cal{B} capable of approximating the linear 2pcf of all possible curved $w$CDM models close to a fiducial flat $\Lambda$CDM model, over a range of separations $r$ relevant for model-agnostic BAO studies. We achieved this using a NN with a custom architecture that we denote \biseq\ (Fig.~\ref{fig:biseq-illustrate}). The discovered basis (Fig.~\ref{fig:biNN2p-basis-residuals}, \emph{left panel}) can approximate \xilin{r} in $w$CDM models with parameters within $\sim5\%$ of their fiducial values to $\simeq0.6\%$ accuracy on average, across the range of length scales explored (Fig.~\ref{fig:biNN2p-basis-residuals}, \emph{right panel}). The corresponding values of scales such as the peak, linear point and zero-crossing of \xilin{r} are also recovered with very high accuracy (Fig.~\ref{fig:biNN2p-scaleresids} and associated discussion). The range of variation of the $w$CDM parameters we allowed above is large compared to current constraints, making our results directly applicable to model-agnostic analyses (see below) in current and upcoming surveys. Below, we briefly summarize some significant aspects of our results and discuss future avenues of exploration.

An important aspect of our analysis was the fact that the training data set used in the \biseq\ discovery of the successful basis \Cal{B} displayed in Fig.~\ref{fig:biNN2p-basis-residuals} was constructed by only varying a small subset $\{\Omega_{\rm m},h\}$ of all the $w$CDM parameters of interest. We demonstrated that this subset is consistent with being the minimal set of parameters needed for producing a sufficiently general basis. In other words, \Cal{B} generalizes to describing \xilin{r} in models for which the weights \Cal{W} (now obtained by least squares fitting rather than the NN itself) can account for variations in many more parameters than used in the training data. Importantly, our results explicitly demonstrate that such a general, model-independent basis indeed exists. Finally, the same results led us to speculate that the same basis \Cal{B} might also generalize to classes of modified gravity models sufficiently close to the fiducial $\Lambda$CDM (section~\ref{subsec:beyondLCDM}), an aspect we will test in the near future.

Our framework is an example of a broader class of problems involving \emph{neural compression} of functions such as \xilin{r;\bt}. This approach has interesting connections to other data compression schemes in the literature, which we discussed in section~\ref{subsec:compare_lit}. As mentioned there, in a future publication we will also report the results of comparing the basis discovered in our approach with those obtained in approaches such as \textsc{moped} or its nonlinear generalizations, when applied in the context of specific models and data sets. Also, while we focused on describing the 2pcf in real space, our approach could also be applied in Fourier space as an alternative to recent analyses such as \citep{aaz21,philcox+21,eggemeier+23,bcv24}.

Our focus in this work was on demonstrating the existence of \Cal{B} and studying its generalization capacity (or degree of completeness) in the space of $w$CDM models. As such, we ignored the problem of using the fitted coefficients (or weights) \Cal{W} -- obtained in some model-agnostic analysis, say -- to \emph{infer} the values of the parameters \bt\ of some specific model. Any such inversion would also have to account for the redshift evolution of the linear 2pcf, which we were able to ignore in this work since this would only introduce a multiplicative scaling of all the weights (through the linear growth factor $D(z)$). Such a scaling, however, will likely lead to important degeneracies in the inferred parameters $\bt(\Cal{W}|z)$. We leave the interesting exercise of calibrating this inversion, which in principle could be achieved using another NN, and quantifying the associated parameter degeneracies in various cosmological models, to future work. 

As mentioned in the Introduction, a primary motivation of our chosen line of approach is to obtain a minimal theoretical basis \Cal{B} for use in model-agnostic analyses of the BAO feature, particularly in redshift space when forward-modelling the 2pcf multipoles $\xi^{(\ell)}(s)$ for $\ell=0,2,4$ as done, for example, by \cite{ps23}. The results presented here will allow us to replace the \emph{ad hoc} choice of monomial basis functions used by \cite{ps23} with the better-motivated basis \Cal{B} discussed above.  We explicitly showed in section~\ref{sec:results} that the monomial basis is far inferior to the \Cal{B} discovered in our approach, leading to errors nearly a factor $\sim20$ larger for the same number of functions. In this context, the authors of \cite{ps23} already noticed that not all of the monomial coefficients allowed to vary in their mock analysis were measured with reasonable statistical significance. Since \Cal{B} is consistent with being a minimal basis, we expect that its use in the context of particular data sets such as DESI LRG, when combined with the approximate maximisation of Bayesian evidence as done by \cite{ps23} and/or the orthogonalization mentioned in section~\ref{subsec:compare_lit}, would lead to a further reduction in the number of varied coefficients and a corresponding improvement in the constraints on cosmologically interesting variables. As noted by \cite{ps23}, however, an important additional ingredient currently missing in such analyses is the inclusion of the effects of scale-dependent bias and mode coupling on the redshift space multipoles $\xi^{(\ell)}(s)$ of the nonlinear 2pcf (see \cite{bkBAO} for a recent study highlighting the importance of scale-dependent bias for the real space nonlinear 2pcf). In a forthcoming publication, we will report the development of such a model, which can then be combined with the basis discovered in this work to set up a complete model-agnostic framework for BAO studies.
 
\section*{Data availability}
Our code implementation of the \biseq\ architecture, along with a simple Jupyter notebook showing how to use an existing instance of \biseq, is publicly available as part of the \texttt{MLFundas} repository at \url{https://github.com/a-paranjape/mlfundas}. The specific example of \biseq\ included there is the same as the one used for generating Figs.~\ref{fig:biNN2p-basis-residuals}-\ref{fig:biNN2p-scaleresids}. A Jupyter notebook with the implementation of training, validation and testing used in this work is available upon reasonable request to AP.

\section*{Acknowledgments}
The research of AP is supported by the Associates Scheme of ICTP, Trieste.  This work made extensive use of the open source computing packages NumPy \citep{vanderwalt-numpy},\footnote{\url{http://www.numpy.org}} SciPy \citep{scipy},\footnote{\url{http://www.scipy.org}} Matplotlib \citep{hunter07_matplotlib},\footnote{\url{https://matplotlib.org/}} and Jupyter Notebook.\footnote{\url{https://jupyter.org}} AP and RKS thank ICTP, Trieste for hospitality during the summer of 2024 when this work was initiated. We thank the anonymous referee for an insightful report.

\bibliography{references}

\appendix
\section{Small-range tests}
\label{app:smallrangetests}
\begin{figure*}[h!]
\centering
\includegraphics[width=0.39\textwidth]{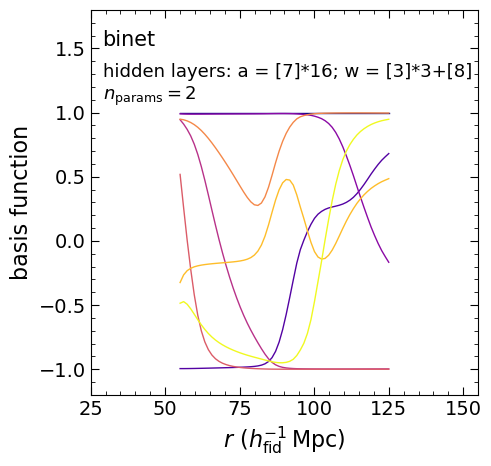}
\includegraphics[width=0.4\textwidth]{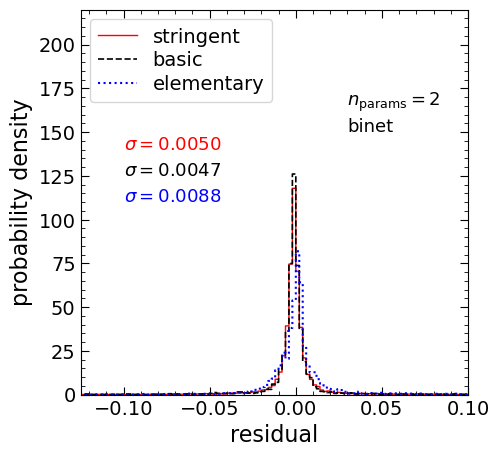}
\caption{Same as Fig.~\ref{fig:biNN2p-basis-residuals}, restricting the range of $r$ for the entire exercise to $55\leq r/(\Mpch) \leq 125$. Note the broader range of the $x$-axis of the \emph{right panel} as compared to the right panel of Fig.~\ref{fig:biNN2p-basis-residuals}.}
\label{fig:biNN2p-smallrange-basis-residuals}
\end{figure*}

To study the impact of the choice of parameters \bt\ varied in the training data, we focus on the smaller range $55\leq r/(\Mpch)\leq 125$, which allows for smaller data sets and network sizes, and correspondingly  smaller compute times that allow us to explore multiple variations. Fig.~\ref{fig:biNN2p-smallrange-basis-residuals} is formatted identically to Fig.~\ref{fig:biNN2p-basis-residuals} and shows the results of training and validation using $\bt=(\Omega_{\rm m},h)$ for this smaller range. In this case, the optimal NN was found to have $L_{\rm b}=16$ and $L_{\rm w}=4$, while the number of basis functions is $n_\ast=8$, i.e., one less than that needed for the broader range in Fig.~\ref{fig:biNN2p-basis-residuals}. The decrease in number of basis functions is sensible considering the reduced variation in the shape of the function. Interestingly, though, the nature of variations in the basis shapes in the \emph{left panels} of Figs.~\ref{fig:biNN2p-basis-residuals} and~\ref{fig:biNN2p-smallrange-basis-residuals} in their common range is rather similar. The \emph{right panel} of Fig.~\ref{fig:biNN2p-smallrange-basis-residuals} shows that the discovered basis is again sufficiently complete to describe the `stringent' test data set (red histogram).

\begin{figure*}[h!]
\centering
\includegraphics[width=0.39\textwidth]{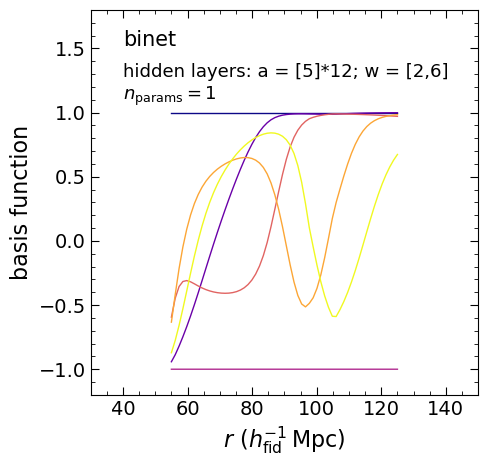}
\includegraphics[width=0.4\textwidth]{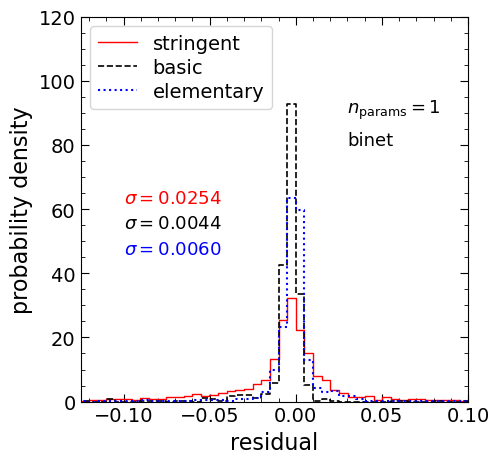}
\caption{Same as Fig.~\ref{fig:biNN2p-smallrange-basis-residuals}, but now varying only $\Omega_{\rm m}$ in the construction of the training and validation data. 
}
\label{fig:biNN1p-smallrange-basis-residuals}
\end{figure*}

\begin{figure*}[h!]
\centering
\includegraphics[width=0.39\textwidth]{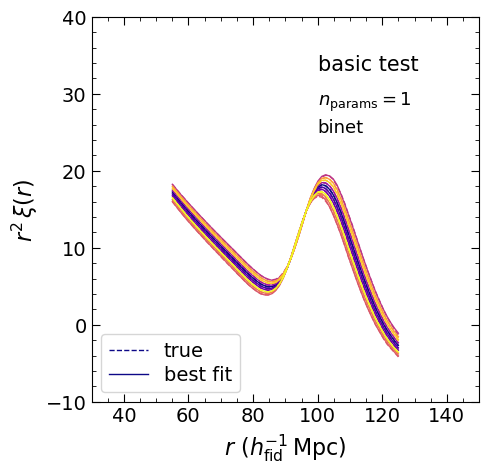}
\includegraphics[width=0.39\textwidth]{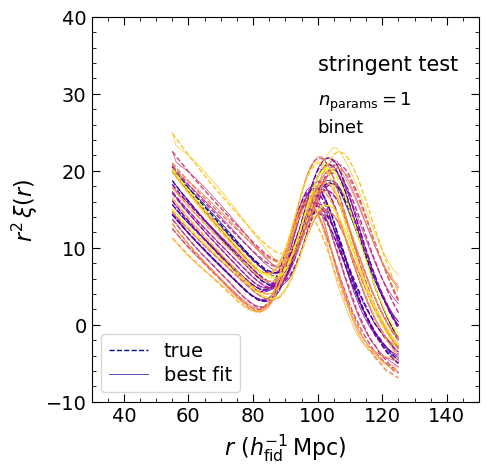}
\caption{Same as Fig.~\ref{fig:biNN2p-tests}, but restricting to the small range of $r$ and varying only $\Omega_{\rm m}$ in the construction of the training and validation data.}
\label{fig:biNN1p-smallrange-tests}
\end{figure*}

We repeated our analysis by varying only $\Omega_{\rm m}$ for the training and validation. In this case, the optimal NN has $L_{\rm b}=12$ and $L_{\rm w}=2$ and requires only $n_\ast=6$ basis functions, shown in the \emph{left panel} of Fig.~\ref{fig:biNN1p-smallrange-basis-residuals}. We remind the reader that the `elementary' test uses weights output by the trained instance of \texttt{NNw}, while the `basic' (and `stringent') test ignores \texttt{NNw} and instead estimates weights using least squares fitting. The \emph{right panel} of Fig.~\ref{fig:biNN1p-smallrange-basis-residuals} shows that, although the `elementary' and `basic' tests are satisfied with similar quality as when varying $\bt=(\Omega_{\rm m},h)$, the `stringent' test when varying only $\Omega_{\rm m}$ for training and validation leads to residuals that are about a factor $4$ larger.  Fig.~\ref{fig:biNN1p-smallrange-tests}, which is formatted identically to Fig.~\ref{fig:biNN2p-tests}, shows why: varying $\Omega_{\rm m}$ alone (cf., the \emph{left panel}) does not produce enough variation in the region around the linear point, and the resulting basis is therefore not versatile enough to describe the effect of varying other parameters (cf., the \emph{right panel}).

\begin{figure*}[h!]
\centering
\includegraphics[width=0.39\textwidth]{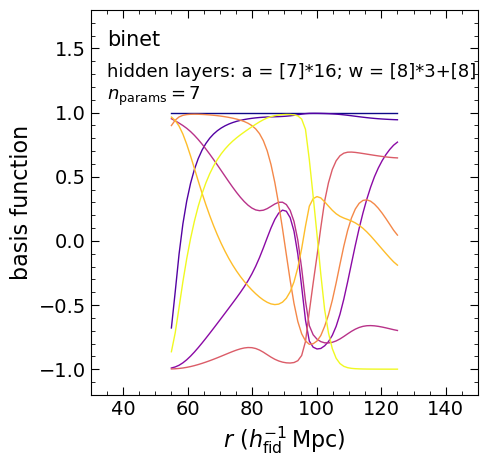}
\includegraphics[width=0.4\textwidth]{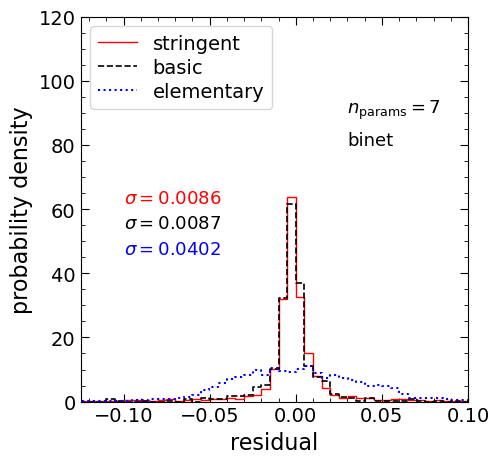}
\caption{Same as Fig.~\ref{fig:biNN2p-smallrange-basis-residuals}, but now varying \emph{all parameters} in the construction of the training and validation data. In this case, the `basic' and `stringent' tests are statistically equivalent to each other.}
\label{fig:biNN7p-smallrange-basis-residuals}
\end{figure*}

We also repeated our analysis by varying \emph{all} 7 parameters for generating the training and validation data sets. In this case, we found the NNs more cumbersome to train and hence stopped the training and validation exercise upon achieving better than $5\%$ validation accuracy. The results are shown in Fig.~\ref{fig:biNN7p-smallrange-basis-residuals}, which is formatted identically to Fig.~\ref{fig:biNN2p-basis-residuals}. The optimal NN now has $L_{\rm b}=16$ and $L_{\rm w}=4$, with $n_\ast=8$ basis functions. Although this is formally the same as the 2-parameter variation case shown in Fig.~\ref{fig:biNN2p-smallrange-basis-residuals}, the larger number of parameters varied in the present case means that the hidden layers of \texttt{NNw} are broader (recall that this width is set to $n_{\rm p}+1$). Consequently, the number of NN weights optimized is also larger ($1134$ for $n_{\rm p}=7$ versus $919$ for $n_{\rm p}=2$). 

Interestingly, despite the fact that \texttt{NNb} has exactly the same architecture for the $\bt=(\Omega_{\rm m},h)$ (i.e., $n_{\rm p}=2$) and $n_{\rm p}=7$ cases, the resulting basis functions show considerably larger short-wavelength variation in the latter case (compare the \emph{left panels} of Fig.~\ref{fig:biNN2p-smallrange-basis-residuals} and Fig.~\ref{fig:biNN7p-smallrange-basis-residuals}, respectively). Alongside, the `stringent' test residuals\footnote{Note that, by construction, the `basic' and `stringent' tests are statistically identical in the $n_{\rm p}=7$ case.} in the $n_{\rm p}=7$ case (\emph{right panel} of Fig.~\ref{fig:biNN7p-smallrange-basis-residuals}) are about a factor $1.7$ larger than those for $\bt=(\Omega_{\rm m},h)$ (\emph{right panel} of Fig.~\ref{fig:biNN2p-smallrange-basis-residuals}). We interpret this as a consequence of not searching a wide enough class of NN architectures to achieve a sub-percent validation accuracy in the $n_{\rm p}=7$ case. Another consequence of this choice is the large difference in magnitude of residuals in the `elementary' and `stringent' (or `basic') tests. It is clear that, in this case, the output of \texttt{NNw} will not match the best-fitting coefficients obtained from the least squares exercise of the `basic' and `stringent' tests. 

Obtaining sufficiently accurate results when varying the full parameter set thus requires considerably more resources than we have used. This will be further compounded with the difficulty in extending the results to the larger range of separations needed for cosmological analyses (see section~\ref{subsec:cosmo}). Overall, we conclude that $\bt=(\Omega_{\rm m},h)$ is consistent with being a minimal set of parameters that need to be varied in order to achieve a sufficiently complete basis \Cal{B} for approximating the linear 2pcf \xilin{r} over the full required range of separations $r$.

\section{Basis function shapes}
\label{app:basisfuncs}
The shapes of the basis functions discovered in our analysis and shown in the \emph{left panel} of Fig.~\ref{fig:biNN2p-basis-residuals} are inherently interesting. The function labelled 2 in Fig.~\ref{fig:biNN2p-basis-residuals}, for example, is evocative of the derivative $\der\xilin{r}/\der r$ with a constant vertical offset \cite[cf. fig.~4 of][]{s-fh21}, while function 9 is similar to the fiducial \xilin{r} itself, but scaled along the horizontal axis. The remaining non-trivial functions show a varying mix of importance between describing the BAO feature versus the broadband shape of the 2pcf.

In the main text, we demonstrated that the basis functions shown in Fig.~\ref{fig:biNN2p-basis-residuals} are sufficiently complete so as to generalize to describing \xilin{r} when varying $7$ $\Lambda$CDM parameters, and speculated as to their generalizability beyond $\Lambda$CDM. On the other hand, one might also ask whether the discovered basis is \emph{over}-complete. In other words, could we have found a basis with $n_* < 8$ that leads to similar performance? The hyperparameter search we reported in section~\ref{subsec:train-val-exc} indicates otherwise; however, that optimization involved a crude exploration of parameter space, limited by computational resources. If an equally acceptable basis set with fewer functions does exist, it would certainly be interesting from the point of view of reducing the dimensionality of parameter inference exercises (cf., e.g., the comparison with the monomial basis reported in section~\ref{sec:results}).

To partially address this question, we consider the set of \emph{weighted} basis functions $B_i(r) \equiv w_i\,b_i(r)$ in the `basic' and `stringent' tests, where the weights $w_i$ are obtained from the respective least squares fitting. The $B_i$ are defined such that their simple sum gives an approximation to the target \xilin{r}. Inspecting the $B_i$ functions can then give a sense for which of the basis functions are most relevant, and whether or not some of them could have been discarded.

\begin{figure*}[h!]
\centering
\includegraphics[width=0.43\textwidth]{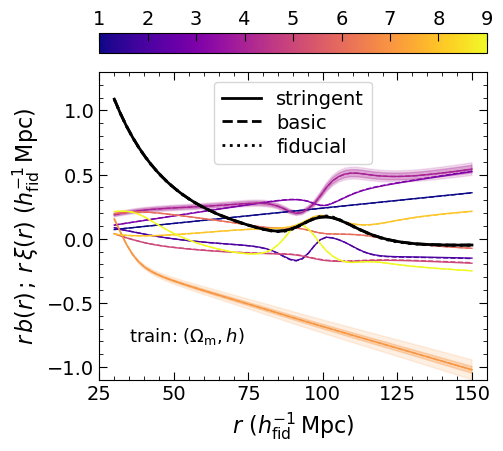}
\includegraphics[width=0.45\textwidth]{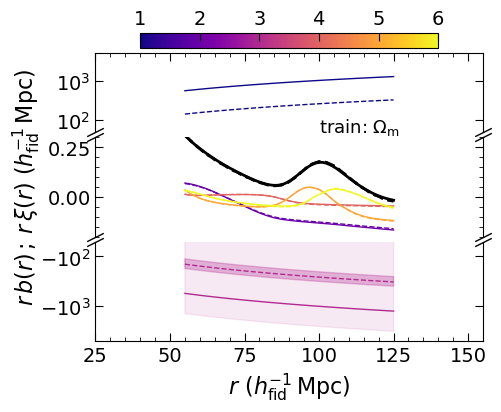}
\caption{Weighted basis functions (coloured lines) $B_i(r) = w_i\,b_i(r)$, with the weights $w_i$ averaged over the `stringent' (solid) and `basic' (dashed) test data sets. The \emph{left panel} uses the basis functions from Fig.~\ref{fig:biNN2p-basis-residuals} (with identical colour-coding), for which the training data varied $\bt=(\Omega_{\rm m},h)$, while the \emph{right panel} uses functions from Fig.~\ref{fig:biNN1p-smallrange-basis-residuals} where the training varied only $\Omega_{\rm m}$. 
Correspondingly formatted black lines show the sum over the $B_i$. Dotted black line shows \xilin{r} in the single, fiducial cosmology. For two representative functions (numbers 4 \& 7 in the \emph{left panel} and numbers 3 \& 6 in the \emph{right panel}), the light (dark) shaded band shows the spread around the mean in the `stringent' (`basic') data sets. For clarity, all functions have been multiplied by $r$. See text for a detailed discussion.}
\label{fig:basisfunc-compare}
\end{figure*}

The \emph{left panel} of Fig.~\ref{fig:basisfunc-compare} shows the \emph{average} of each $B_i$ over all parameter vectors comprising the `stringent' (solid) and `basic' (dashed) data sets; these are nearly indistinguishable. The colours are assigned as in Fig.~\ref{fig:biNN2p-basis-residuals} and indicated by the colour bar. The black solid and dashed lines show the respective sums over the $B_i$. These are also nearly indistinguishable, which is not surprising given the behaviour of their component functions (also recall from the \emph{right panel} of Fig.~\ref{fig:biNN2p-basis-residuals} that the `basic' and `stringent' tests both led to sub-percent accuracy for this configuration). Interestingly, the black solid and dashed lines are also nearly indistinguishable from the black dotted line, which shows \xilin{r} for the \emph{single} fiducial cosmology, indicating that $\sim5\%$ variations of the cosmological parameters are approximately linear in the parameters. Overall, we see comparable contributions from each of the basis functions; there are clear cancellations at large $r$ and collective enhancements at small $r$. It is therefore difficult to isolate any particular basis function that contributes the most or least to describing \xilin{r}, supporting our argument against over-completeness. For reference, we also show the standard deviation around the mean $B_i$ for two representative functions, with the light (dark) bands showing the width for the stringent (basic) test data. As expected, the stringent data have a larger width (since a larger number of parameters were varied), but both widths are relatively small. We have checked that the other functions behave qualitatively similarly.

The \emph{right panel} of Fig.~\ref{fig:basisfunc-compare} is formatted and constructed similarly to the \emph{left panel}, and shows results for the 6 basis functions displayed in Fig.~\ref{fig:biNN1p-smallrange-basis-residuals} where only $\Omega_{\rm m}$ was varied in generating the training data. Contrary to the \emph{left panel}, we now notice that basis functions 1 and 3 have extremely large contributions that are nearly an order of magnitude different between the `basic' and `stringent' data sets, but which nevertheless nearly precisely cancel in each data set. This is also evident from the \emph{left panel} Fig.~\ref{fig:biNN1p-smallrange-basis-residuals}, which shows that both of these are essentially constant (function 1 by construction and 3 due to the NN training) and hence perfectly degenerate with each other.
This means the final shape of the target output is largely determined by the other basis functions, whose contributions are comparable to each other and to the final output (and are also very similar between the two data sets). Only the overall vertical offset is influenced by the degenerate combination of functions 1 and 3. Correspondingly, we also see that the spread in the contribution of function 3, especially in the `stringent' data set, is very large compared to, say, that of function 6 whose spread is displayed but is smaller than the line thickness.\footnote{We caution against over-interpreting the absolute magnitude of the spreads displayed in both panels, since these are entirely determined by our assignment of a constant \emph{ad hoc} error to the target output.} In this case, a basis set with 5 functions (say, excluding function 3) would have produced identical results for the target.

Finally, returning to the example of the primary analysis (Figs.~\ref{fig:biNN2p-basis-residuals}, \ref{fig:biNN2p-tests} and the \emph{left panel} of Fig.~\ref{fig:basisfunc-compare}) we note that the basis functions 3, 8 and 9 all show a BAO peak-like feature, but shifted horizontally, while the functions 2 and 4 show (inverted) derivative-like features, also shifted horizontally in terms of the locations of the extrema. Of course, the broadband nature of these functions is quite different from one another, but our focus is on the scales near the BAO feature. These horizontal shifts are interesting considering that the dominant effect of varying the dimensionless Hubble constant $h$ is to produce a horizontal scaling of \xilin{r} when $r$ is measured in the fixed units of Mpc or $h_{\rm fid}^{-1}$Mpc. (There are also additional effects of varying $h$: since we hold $\Omega_{\rm m},\Omega_{\rm b}$ fixed while doing so, varying $h$ also varies the physical densities $\Omega_{\rm m}h^2,\Omega_{\rm b}h^2$.) We believe the shifted features in the different basis functions are responsible for accounting for this scaling behaviour, which is consistent with the fact that the spread in locations of the features is comparable to the $5\%$ variation of $h$ in the test data sets. Since the scaling is, in some sense, a trivial effect, this means our choice of architecture and input features has forced the NN to `waste' some effort in learning this pattern. It will be very interesting to revisit our architectural choices so as to explicitly account for the scaling property of $h$ during training. (Similar comments apply to the scalings suggested by the widely used parameterizations of the power spectrum which appear in Appendix G of \cite{bbks86}.)  Since we have already seen that variations of $\{\Omega_{\rm m},h\}$ are sufficient to produce a complete basis, this would mean that a network architecture which explicitly accounts for scaling of $r$ might produce highly accurate results by training on $\Omega_{\rm m}$ variations alone. Although outside the scope of the present work, we will return to this issue in the future.

\end{document}